\documentclass[Afour,sagev,times]{sagej}

\usepackage{moreverb,url}

\usepackage{xcolor}
\usepackage{subcaption}
\usepackage{float}
\usepackage{tabularx}
\usepackage{booktabs}
\usepackage{xcolor}
\usepackage[linesnumbered,ruled,vlined]{algorithm2e}
\usepackage{amsmath}
\usepackage[most]{tcolorbox}
\tcbuselibrary{skins, breakable, theorems}
\usepackage{booktabs}   
\usepackage{tabularx}  
\usepackage{multirow}   
\usepackage{array}     
\newcommand\BibTeX{{\rmfamily B\kern-.05em \textsc{i\kern-.025em b}\kern-.08em
T\kern-.1667em\lower.7ex\hbox{E}\kern-.125emX}}

\def\volumeyear{2026}

\definecolor{review}{rgb}{0, 0, 0}
\definecolor{review2}{rgb}{0, 0, 0}

\begin{document}
\title{Toward Reliable Scientific Visualization Pipeline Construction with Structure-Aware Retrieval-Augmented LLMs}

\author{Guanghui Zhao\affilnum{1,2,$^{*}$}, Zhe Wang\affilnum{2,3,$^{*}$}, Yu Dong\affilnum{3}, Guan Li\affilnum{2,3}, GuiHua Shan\affilnum{1,2,3,$^\dagger$}}

\affiliation{\affilnum{1}Hangzhou Institute for Advanced Study, University of Chinese Academy of Sciences\\
\affilnum{2}University of Chinese Academy of Sciences \\
\affilnum{3}Computer Network Information Center, Chinese Academy of Sciences\\
\affilnum{$^\dagger$}Guihua Shan is the corresponding author. E-mail: sgh@cnic.cn\\
\affilnum{$^*$}These authors contributed equally to this work.}

\begin{abstract}

Scientific visualization pipelines encode domain-specific procedural knowledge with strict execution dependencies, making their construction sensitive to missing stages, incorrect operator usage, or improper ordering. Thus, generating executable scientific visualization pipelines from natural-language descriptions remains challenging for large language models, particularly in web-based environments where visualization authoring relies on explicit code-level pipeline assembly. In this work, we investigate the reliability of LLM-based scientific visualization pipeline generation, focusing on vtk.js as a representative web-based visualization library. We propose a structure-aware retrieval-augmented generation workflow that provides pipeline-aligned vtk.js code examples as contextual guidance, supporting correct module selection, parameter configuration, and execution order. We evaluate the proposed workflow across multiple multi-stage scientific visualization tasks and LLMs, measuring reliability in terms of pipeline executability and human correction effort. To this end, we introduce correction cost as metric for the amount of manual intervention required to obtain a valid pipeline. Our results show that structured, domain-specific context substantially improves pipeline executability and reduces correction cost. We additionally provide an interactive analysis interface to support human-in-the-loop inspection and systematic evaluation of generated visualization pipelines.

\end{abstract}

\keywords{Scientific visualization, Retrieval-Augmented Generation (RAG), Code generation, Human-in-the-loop evaluation, Error correction}

\maketitle


\section{Introduction}

Scientific visualization pipelines encode domain-specific procedural knowledge through a sequence of tightly coupled operations that transform raw scientific data into visual representations.
Such pipelines typically consist of multiple stages, including data loading, transformation, and rendering, where correctness depends not only on individual operator implementations, but also on their parameter semantics, execution order, and inter-stage dependencies.
In these pipeline-oriented settings, errors often propagate across stages and invalidate the execution of the entire workflow, making successful end-to-end execution a fundamental requirement for correctness.
Consequently, constructing valid scientific visualization pipelines requires precise coordination of structure and semantics, rather than merely producing visually plausible output.


Recent systems improve LLM-based scientific visualization pipeline construction either by closing the loop with execution feedback (e.g., iterative error correction)~\cite{chatvis} or by augmenting generation with retrieved examples~\cite{11279991}. 
Most of these efforts target VTK~\cite{2006vtk}/Python-style visualization scripting generation, where abundant public code and documentation increase the likelihood that foundation models have encountered relevant patterns during pretraining, making one-shot executable generation more attainable. 
However, in domain-specific settings with scarce in-domain corpora and evolving toolchains, LLMs become more prone to hallucinated APIs, missing pipeline stages, and invalid parameters, and trustworthy end-to-end scientific visualization pipeline generation remains challenging.

The vtk.js inherits VTK-style pipeline abstractions, yet its web-centric ecosystem is smaller and evolves more rapidly, leaving its APIs and usage patterns less represented in foundation-model pretraining and thereby increasing the risk of hallucinated calls and invalid parameterizations. 
Meanwhile, the browser runtime is lightweight and reproducible, enabling automated execution-based testing and cross-backend replication of generated pipelines. 
Moreover, modern web visualization frameworks such as the \texttt{trame} ecosystem built on vtk.js~\cite{jourdain2025trame} have lowered the barrier for scientific use cases and accelerated adoption in domains such as biological data analysis and microscopy image exploration~\cite{bohak2020web,gupta2022interactive}. 
We therefore select vtk.js in a web-based environment as our evaluation testbed for investigating reliable scientific visualization pipeline construction.

In this work, we conceptualize scientific visualization pipeline generation as a reliability problem rather than a visual quality problem.
Unlike chart-oriented visualization, where partial correctness may still yield interpretable output, scientific visualization pipelines require all essential stages to be present and correctly ordered in order to execute. Consequently, reliability in this context is determined by pipeline completeness, structural validity, and executability, rather than aesthetic or perceptual criteria. Focusing on scientific visualization with vtk.js as a representative domain-specific library, we conduct a retrieval-augmented workflow that explicitly encodes pipeline structure, module compatibility, and execution order during context construction. To support iterative exploration and validation, we further provide a lightweight front-end interface that keeps humans in the loop: users can inspect retrieved evidence and generated pipeline steps, execute the pipeline, and provide targeted feedback for refinement. We evaluate our approach across a set of multi-stage scientific visualization tasks and multiple LLMs, using executability and correction cost as measures of generation reliability. Based on these evaluations, we derive several key contributions:

\begin{itemize}
    \item We empirically examine the reliability of generating executable scientific visualization pipelines from high-level task descriptions, showing that unstructured generation frequently produces non-executable workflows.
    \item We demonstrate that incorporating structure-aware, pipeline-aligned contextual examples significantly improves pipeline executability and reduces manual correction cost in web-based scientific visualization.
    \item We examine recurring failure cases in scientific visualization pipeline construction, highlighting structural and semantic challenges—such as derived-field computation and parameter constraints—that remain difficult to address through retrieval alone.
\end{itemize}

\section{Related Work}

Generating visualization results from natural-language descriptions has been an active research topic for many years, evolving from early symbolic NLP approaches to recent deep learning–based methods~\cite{10121440}.
Within this broad area, automated code generation plays a central role in bridging user intent and executable visualization results.
We refer readers to recent surveys for a comprehensive overview of large language model (LLM)–based code generation, including data curation strategies, retrieval-augmented generation, and evaluation protocols~\cite{jiang2024survey}.
In this section, we review prior work most relevant to our study and position our contribution with respect to how these approaches support visualization pipeline construction and executability.

\paragraph{LLM-assisted visualization generation.}\textcolor{review2}{Several studies have explored the use of LLMs for automating information visualization, including grammar-agnostic chart generation~\cite{dibia2023lida,10121440}, autonomous workflows~\cite{zhang2023data}, and human-in-the-loop authoring~\cite{wang2023data}. In contrast, other systems focus on specific scripting environments.}
ChatVis~\cite{chatvis} presents an iterative assistant that generates Python scripts for data analysis and visualization using LLMs.
AuraGenome~\cite{zhang2025auragenome} similarly supports progressive generation of JavaScript code for genomic visualization through stepwise refinement.
Both systems employ chain-of-thought prompting to decompose user requests and rely on iterative correction to improve generated code.
These studies demonstrate the effectiveness of LLM assistance for visualization scripting; however, they primarily depend on few-shot prompting and iterative refinement, without explicitly incorporating retrieval from a structured, domain-specific code corpus.

VisGene~\cite{11279991} proposes a self-improving, multi-agent framework for scientific visualization in high-performance computing environments.
The system integrates LLM-generated VTK/Python modules with pre-existing visualization tools and employs retrieval-augmented generation alongside fine-tuned vision models for feature-based visual question answering.
While VisGene improves robustness through agent orchestration and model adaptation, it focuses on desktop-oriented scientific visualization workflows.
In contrast, our work targets web-based scientific visualization using vtk.js, where sparse documentation and rapidly evolving APIs pose distinct challenges for generating executable visualization pipelines.
Moreover, we explicitly quantify the human editing effort required to correct generated code through a correction cost metric, rather than relying solely on iterative refinement.

\paragraph{Natural language to visualization systems.}
\textcolor{review2}{Several studies have explored natural-language interfaces for visualization generation, particularly for chart-centric scenarios~\cite{dibia2019data2vis, chen2021plotcoder, luo2021natural, han2023chartllama}.}
MatPlotAgent~\cite{yang-etal-2024-matplotagent} introduces a multi-agent framework that automates two-dimensional scientific chart generation and presents MatPlotBench, a benchmark consisting of human-verified test cases.
Similarly, Chat2VIS~\cite{10121440} surveys progress in natural language to visualization (NL2VIS) and proposes a general framework for transforming user queries into chart visualizations with associated evaluation protocols.
These systems primarily emphasize chart specification and visual encoding correctness.
By contrast, our work focuses on three-dimensional scientific visualization pipelines, where correctness is determined by successful end-to-end execution rather than solely by visual plausibility.

\paragraph{Retrieval-augmented code generation.}
Retrieval-augmented code generation has been studied in broader programming contexts as a means to mitigate hallucinations and improve domain specificity, \textcolor{review2}{as demonstrated by RepoCoder~\cite{zhang2023repocoder} and DocPrompting~\cite{zhou2022docprompting}. Furthermore, handling complex libraries with strict module dependencies often requires specialized context augmentation to ensure proper API selection and execution ordering~\cite{qin2023toolllm}, while execution-centric benchmarks have proven essential for evaluating real-world code validity beyond static analysis~\cite{yang2023intercode}.} Koziolek et al.~\cite{koziolek2024llm} present an LLM-based retrieval-augmented workflow for control system programming, combining document loading, chunking, embedding, and vector-based retrieval to guide code synthesis from natural-language descriptions.
While our workflow shares conceptual similarities with this approach, our focus differs in two key aspects.
First, we investigate retrieval-augmented generation in the context of scientific visualization pipelines, which impose strict structural and execution constraints.
Second, we construct a dedicated benchmark with segmented vtk.js code blocks and evaluate generation quality through systematic, task-oriented experiments rather than general-purpose programming tasks.

Overall, prior work has demonstrated the promise of LLMs for visualization authoring and code generation.
However, existing approaches either focus on chart-level visualization or rely on iterative correction and agent orchestration, without explicitly examining how retrieval and structured contextual information affect the executability of scientific visualization pipelines.
Our work addresses this gap by investigating structure-aware retrieval for web-based scientific visualization using vtk.js.

\section{Background and Motivation}

We focus on two guiding questions: why web-based scientific visualization has become an important setting for visualization authoring, and why generating executable visualization pipelines in this context presents unique challenges for large language models (LLMs).
We first discuss the shift from desktop-oriented scientific visualization systems to web-based solutions, and then use a motivating example to illustrate the limitations of off-the-shelf LLMs and the role of structured contextual knowledge.

\subsection{From Desktop-Centered to Web-Based Scientific Visualization}

Traditional scientific visualization systems, such as those built upon VTK and ParaView~\cite{ayachit2015paraview}, have long provided powerful capabilities for constructing complex visualization pipelines.
These systems primarily target expert users and rely on interactive graphical interfaces or low-level scripting to manually assemble and refine visualization workflows.

In recent years, scientific visualization has increasingly moved toward web-based environments.
This shift is driven not merely by technological trends, but by practical requirements such as cross-platform accessibility, zero-install deployment, remote collaboration, and tight integration with web-based data services and analysis pipelines~\cite{kitware_webgpu,vtk_js}.
Web-based scientific visualization systems are often used in exploratory and collaborative settings, where users may not possess detailed knowledge of underlying visualization libraries.
In these scenarios, high-level, intent-driven interaction becomes a natural interface paradigm, increasing the demand for automatic or semi-automatic visualization pipeline generation.

Vtk.js exemplifies this transition by providing a web-native library for interactive three-dimensional scientific visualization directly in the browser.
It has been adopted in production systems across domains such as volumetric and medical imaging~\cite{volview_docs,ohif_docs_v2}, and integrates naturally with modern front-end frameworks and toolchains~\cite{vtk_js}.
However, from the perspective of LLM-based code generation, vtk.js presents a challenging target.
Compared to the extensive corpus of desktop-oriented VTK examples and documentation~\cite{vtk_examples}, publicly available vtk.js code is relatively sparse, and its APIs continue to evolve.
As a result, even large pretrained LLMs often struggle to generate correct and executable vtk.js visualization pipelines from high-level natural-language descriptions.

\begin{figure*}[t]
\centering
\includegraphics[width=1\linewidth]{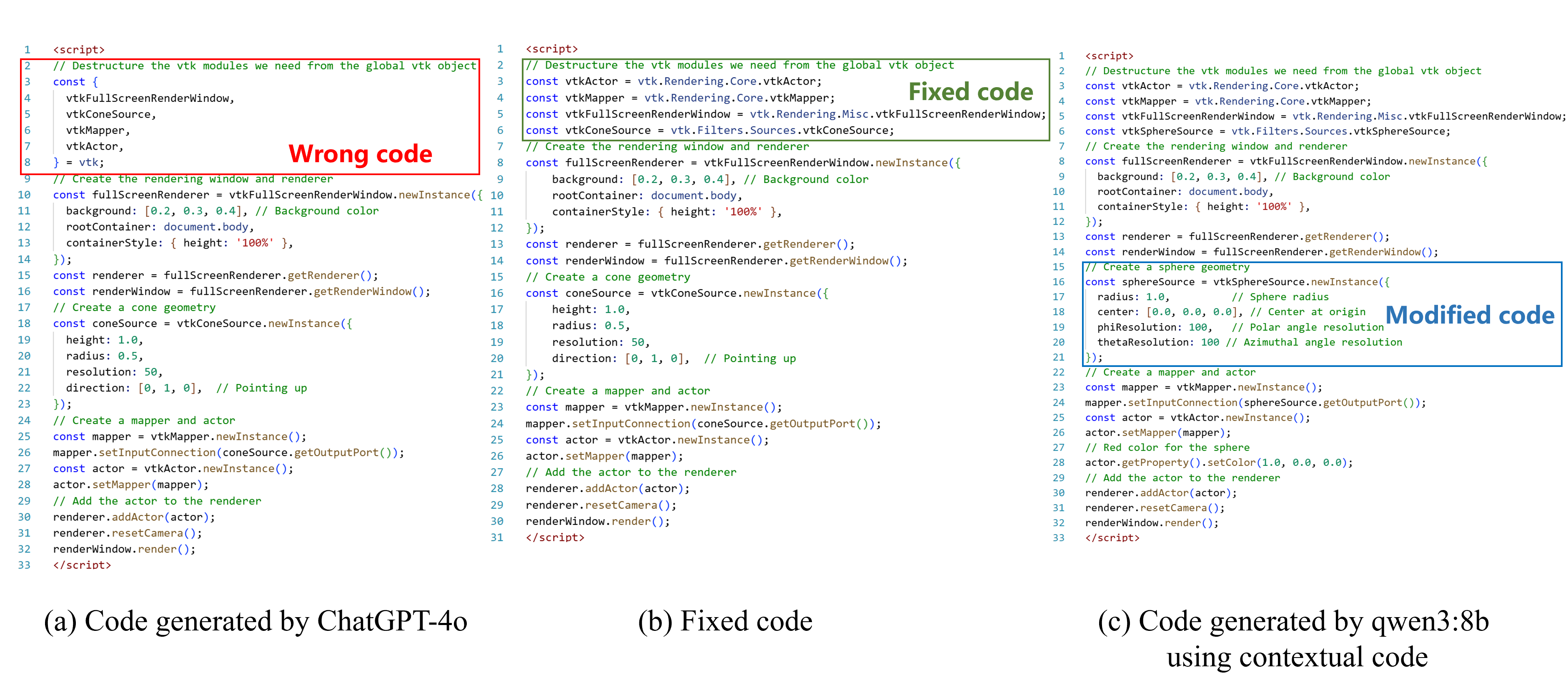}
\caption{(a) shows raw code generated by ChatGPT-4o, (b) shows the fixed code based on raw code. (c) shows results using fixed code as context information and let qwen3:8b to create a sphere source instead of cone. We only show vtk.js code.
}

\label{fg:mtv:exp1}
\end{figure*}

\subsection{Motivating Example: Failures in Web-Based Pipeline Generation}

To illustrate these challenges, we consider a simple but representative example in which an LLM is prompted to generate an HTML page that renders a basic geometric object using vtk.js.
Figure~\ref{fg:mtv:exp1}(a) shows the JavaScript code produced by ChatGPT-4o in response to this prompt.
Despite the simplicity of the task, the generated code omits essential stages of the vtk.js rendering pipeline, resulting in a non-executable program.

Figure~\ref{fg:mtv:exp1}(b) shows a manually corrected version of the code that restores the missing pipeline components.
When the corrected example is used as context, and a similar prompt is given—such as requesting a different geometric source—the open-source qwen3:8B model~\cite{yang2025qwen3} generates correct vtk.js code in a single pass, as shown in Figure~\ref{fg:mtv:exp1}(c).
Notably, this smaller model succeeds once it is conditioned on a high-quality, domain-specific example, despite having substantially fewer parameters than ChatGPT-4o ~\cite{openai2024gpt4technicalreport}.

This example highlights a key observation: failures in web-based scientific visualization code generation are not solely attributable to model capacity.
Instead, failures stem from a lack of explicit access to the structural and semantic knowledge that governs scientific visualization pipelines.
Without such context, LLMs are prone to hallucinations, missing pipeline stages, and incorrect operator usage, and correcting these errors requires significant domain expertise.
Conversely, when relevant pipeline examples are available as contextual guidance, even small models can generate executable visualization pipelines for complex tasks.

\subsection{Problem Context and Research Questions}
The motivating example suggests that improving automatic generation of web-based scientific visualization pipelines requires mechanisms that expose LLMs to structured, domain-specific knowledge at inference time. Retrieval-augmented generation offers a promising direction by enabling models to condition on curated examples that encode pipeline structure, operator compatibility, and execution order. Motivated by these observations, we investigate the factors that influence the reliability of scientific visualization pipeline generation in web-based settings, including the role of structured contextual information and human correction. Specifically, we pose the following research questions:

\textbf{RQ1.} What types of structural and semantic errors most commonly prevent LLM-generated scientific visualization pipelines from being executable in web-based settings?

\textbf{RQ2.} To what extent can structured, domain-specific contextual information improve pipeline executability, and how does this improvement vary across visualization tasks and language models?

\textbf{RQ3.} How should human correction effort be systematically characterized to more faithfully reflect the reliability of generated scientific visualization pipelines beyond binary success or failure?


\section{Method}

\begin{figure}[H]
    \centering 
    \includegraphics[width=0.45\textwidth]{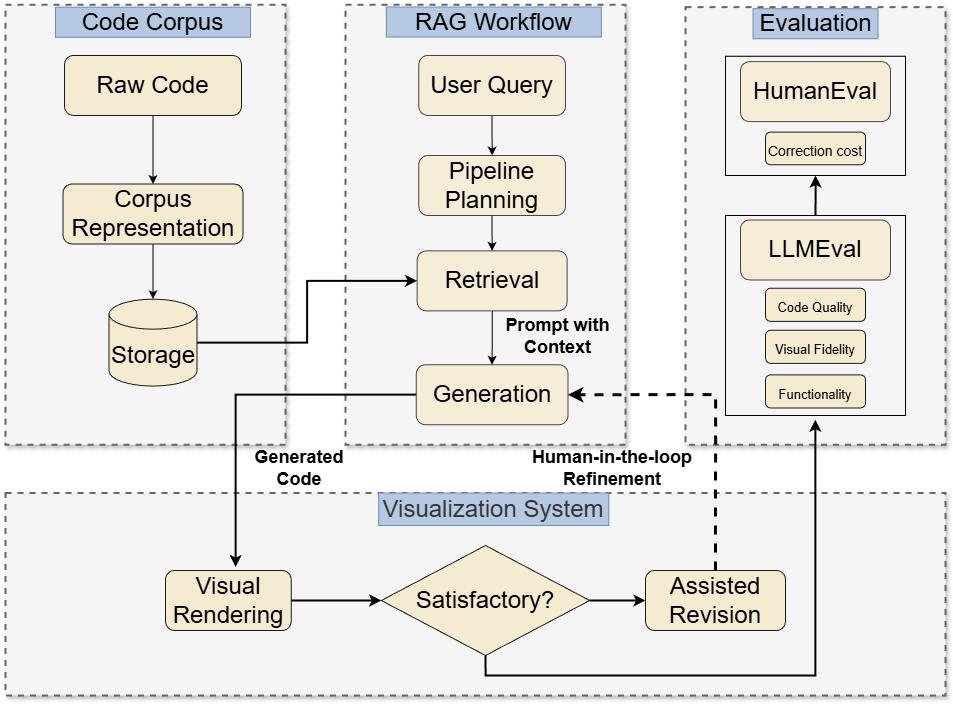}  
    \caption{\textcolor{review}{The workflow presented in this paper includes four key modules: code corpus preparation, RAG workflow, result evaluation and visualization system.}}
    \label{fig:framework}
\end{figure}

To address reliability challenges discussed above, we formulate web-based scientific visualization code generation as a pipeline construction problem with strict structural and execution constraints.
Rather than treating code generation as free-form synthesis, our research emphasizes the preservation of pipeline stages, module compatibility, and execution order required by vtk.js.
We first design a RAG-powered workflow for web-based scientific visualization, which integrates data preparation, retrieval-augmented generation, evaluation principles, and visualization. Then we use the presented workflow to evaluate LLMs in different scientific visualization pipeline construction scenarios.

Figure~\ref{fig:framework} illustrates the overall architecture. 
\textcolor{review}{The workflow begins with the construction of a curated Code Corpus, which feeds directly into the RAG Workflow. Here, the system retrieves structure-aware examples to guide the generation of executable code.
Crucially, the process incorporates a Human-in-the-loop mechanism within the Visualization System: a decision node determines if the rendering is satisfactory. If not, an iterative feedback loop (dashed line) triggers assisted revision. Once the pipeline is validated, it passes to the Evaluation stage, which serves as a downstream dependency to quantify reliability through both automated LLMEval and HumanEval .}

\subsection{Corpus Preparation}
\label{method:codec}

The purpose of corpus preparation is not merely to collect example code, but to construct a structure-addressable original corpus that supports module-level retrieval and pipeline-aligned reasoning.

To construct the original corpus, we first conducted a comprehensive review of the vtk.js syntax documentation and official examples. However, the official vtk.js examples cannot be directly adopted, as they lack metadata extraction and annotation, and do not contain the corresponding prompts required for our code generation tasks. To address this limitation, we carefully selected and refined examples that include well-defined prompts and are implemented purely in HTML and JavaScript, without any additional or complex dependencies. 

We categorized the tasks into three groups: I/O, filtering, and rendering. We curated representative tasks from official examples and common usage patterns to cover key stages of a typical visualization pipeline, including data loading, processing, rendering, and interaction. These tasks serve as reusable building blocks and external knowledge for constructing complex visualization workflows. All examples can be executed and validated directly in a standard web browser, ensuring high accessibility and reproducibility.
\textcolor{review}{In addition to the source code, we provide a description file that outlines the overall goal of the code snippet, as well as a meta file that supports indexing and matching within the corpus.}

\begin{figure}[ht]
    \centering 
    \includegraphics[width=0.46\textwidth]{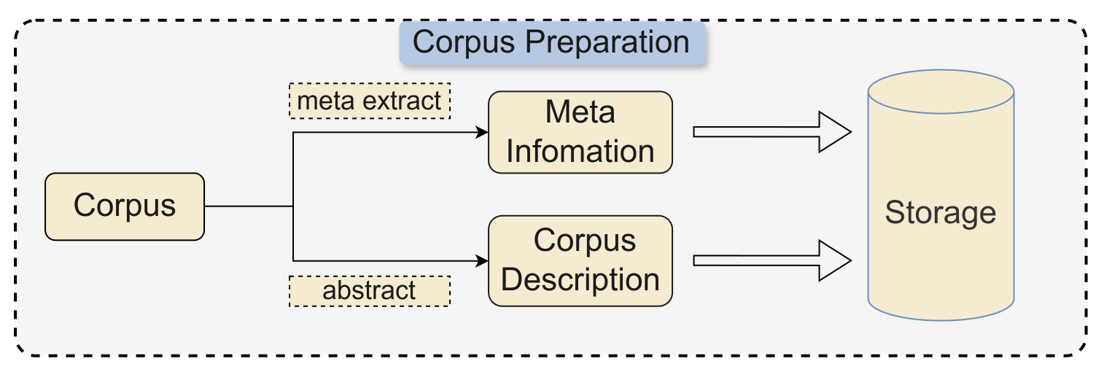}  
    \caption{Overview of the Corpus Preparation}
    \label{fig:corpus}
\end{figure}

To better support the RAG system and highlight the core characteristics of the corpus, we performed additional processing on the original datasets. This included metadata extraction and module indexing, which together enhance the corpus's structural organization and machine readability. Figure~\ref{fig:corpus} details the main steps of corpus preparation.

\subsubsection{Metadata Extraction}
\label{sec:metaextraction}

To support the effective retrieval of visualization code, we developed a pipeline for extracting structured metadata from our corpus. This process generates a standardized, machine-readable representation that captures the essential characteristics of each code sample, which is foundational for subsequent semantic analysis and code generation tasks.

The extraction pipeline operates sequentially through three key stages. It begins with file-level metadata extraction, which identifies fundamental attributes such as the file path, target visualization task, and core functional modules.Next, the module identification stage analyzes the code to catalog the specific VTK.js components utilized; this output is directly used to compute module overlap in the re-ranking stage during code retrieval. Finally, all extracted information is structured and serialized into a consistent JSON format. The resulting metadata, particularly the descriptive elements, thereby enables efficient and structure-aware code retrieval.

\subsection{Retrieval-Augmented code generation workflow}
\label{method:racg}

\begin{figure}[H]
    \centering 
    \includegraphics[width=0.48\textwidth]{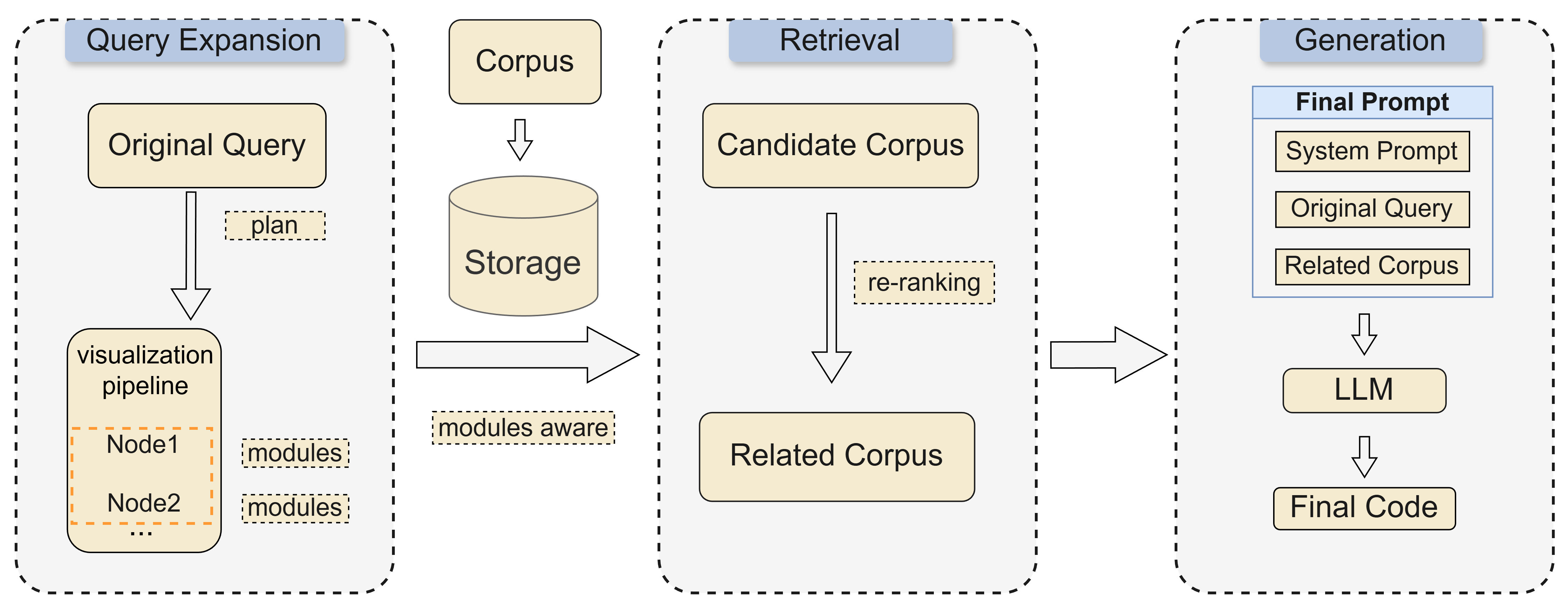}  
    \caption{Details of the Structure-Aware Retrieval-Augmented Generation (RAG) workflow
    }
    \label{fig:racg}
\end{figure}

The RAG workflow proposed in this work transforms ambiguous user intent into executable vtk.js scripts through three coordinated stages: Pipeline Planning, Retriever, and Generation. As illustrated in Figure~\ref{fig:racg}, the user's original intent is first decomposed into a structured pipeline consisting of discrete pipeline nodes. During the matching stage, the system identifies and selects validated \textcolor{review}{corpora}. Crucially, this selection is based on modules of planned pipeline nodes. The retrieved content is then incorporated into the final prompt—together with the initial query and system instructions—as few-shot examples to enhance the LLM’s ability to produce high-quality visualizations. The details of each stage are described below.

\subsubsection{Pipeline Planning}

First, an initial user intent is provided as input. Unlike simple code generation tasks, constructing a scientific visualization workflow requires a \textcolor{review}{strict} pipeline of operations—data loading, processing, and rendering. 
Direct generation from natural language often leads to structural errors (e.g., missing filters or incorrect connections). Therefore, instead of directly retrieving code, we first decompose the user’s intent into a logical Pipeline Plan. This step bridges the gap between ambiguous natural language and the procedural logic required by vtk.js."

The planning stage is executed by an Plan LLM, which transforms the user's unstructured request into a structured sequence of discrete pipeline nodes. Each node represents a specific functional unit within the visualization graph. The architect outputs following schema:

\begin{itemize}
    \item \textbf{\texttt{phase}}: Specifies the logical execution stage in the pipeline (e.g., Data Loading, Filtering, Rendering).
    \item \textbf{\texttt{name}}: The specific functional task identifier.
    \item \textbf{\texttt{modules}}: A list of Candidate Modules of vtk.js predicted to implement this node. These serve as structural constraints for downstream retrieval.
    \item \textbf{\texttt{weight}}: \textcolor{review}{An integer representing the Node Weight, determining the priority of this node during the retrieval matching process.}
    \item \textbf{\texttt{description}}: A detailed natural-language instruction used for subsequent instantiation.
\end{itemize}

\textcolor{review}{To achieve this effect, we assign weights based on the closeness of each pipeline node to the user's primary intent, explicitly distinguishing between three structural layers. 
First, core process ($w \in [8, 10]$) encode the essential logic required to fulfill the user's specific request. For instance, in slicing tasks, this includes filters such as \texttt{vtkImageSlice}; for volume rendering tasks, this category extends to visual mappings like \texttt{vtkVolumeProperty}. 
Second, contextual configuration ($w \in [5, 7]$) steps support the core task but are secondary to the primary objective. This typically involves adjusting presentation attributes via \texttt{vtkCamera} (e.g., \texttt{setPosition}, \texttt{setViewUp}), modifying actor appearance through \texttt{vtkProperty} (e.g., \texttt{setAmbient}, \texttt{setDiffuse}), or applying standard scalar mappings that do not alter data topology using \texttt{vtkColorTransferFunction} or \texttt{vtkLookupTable}. 
Finally, scene management ($w \in [1, 4]$) covers standard boilerplate code required to initialize the visualization environment and coordinate rendering resources.}


The following example shows how a slice-related prompt is expanded into a structured pipeline using qwen3-32b!\cite{yang2025qwen3}. A representative user query is as follows:

\begin{tcolorbox}[
    colback=gray!10,       
    colframe=black,        
    title=User Query,   
    fonttitle=\bfseries\small, 
    coltitle=white,        
    arc=1mm,
    width=24em,
    boxrule=0.3pt
]
Generate HTML with vtk.js to visualize sliced rotor data by loading the dataset from [data server], setting ``Pressure'' as the active scalar. Applying a Y-axis slice at 80\% depth, mapping pressure values to a blue $\rightarrow$ white $\rightarrow$ red color gradient with full opacity, Adding an XYZ axis orientation marker for spatial reference, and no GUI controls.
\end{tcolorbox}

\textcolor{review}{The pipeline generated from the user query above is listed as follows:}

\noindent

\begin{tcolorbox}[
    colback=gray!10,       
    colframe=black,        
    title=Pipeline Plan,   
    fonttitle=\bfseries\small, 
    coltitle=white,        
    arc=1mm,
    width=24em,
    boxrule=0.3pt
]

$\bullet$ \textbf{Data Loading} ($w=3$): Load the rotor VTI dataset from [data server] using \texttt{vtkXMLImageDataReader} and \texttt{vtkHttpDataSetReader}.

$\bullet$ \textbf{Data Processing} ($w=10$): Set active scalar to ``Pressure'' and apply a Y-axis slice at 80\% depth using \texttt{vtkImageSlice} and \texttt{vtkPlane}.

$\bullet$ \textbf{Visualization Setup} ($w=6$): Map ``Pressure'' values to a Blue-White-Red gradient with full opacity using \texttt{vtkColorTransferFunction}.

$\bullet$ \textbf{UI Components} ($w=2$): Add an orientation marker (XYZ axes) to the scene for spatial reference via \texttt{vtkOrientationMarkerWidget}.

$\bullet$ \textbf{Rendering} ($w=2$): Finalize the visualization pipeline using \texttt{vtkGenericRenderWindow} and \texttt{vtkRenderWindowInteractor}.
\end{tcolorbox}\textbf{}

This example demonstrates that the expander effectively transforms user intent into well-structured and semantically rigorous sub-tasks. The pipeline provides clearer, more detailed instructions for each stage of the workflow—including data loading, processing, visualization, UI configuration, and rendering—resulting in a more interpretable and actionable specification during code generation.


\begin{algorithm}[ht]
\small
\DontPrintSemicolon

\SetKwInOut{Input}{Input}
\SetKwInOut{Output}{Output}

\SetKwFunction{Retrieve}{RetrieveByModules}
\SetKwFunction{CountHits}{CountHits}
\SetKwFunction{Normalize}{NormalizeWeights}
\SetKwFunction{Merge}{Merge}
\SetKwFunction{Sort}{Sort}

\caption{\textcolor{review}{Pseudocode of corpus example retrieval}}
\label{alg:vtksercher_v2_simple}

\Input{
\BlankLine
    $D$: List of $(M, \text{code}, \text{des})$\;

    $Q$: \textcolor{review}{List of $(M_{node}, w_{node})$}\; 
    $K$: the number of corpus examples to return 
}
\Output{$R$: List of Top-$K$ corpus examples}

\BlankLine 
\tcp{\textcolor{review}{Auxiliary data structures initialized within the algorithm}}

$C$: Empty list of $(M, \text{code}, \text{score})$

$W_{list} \gets$ \Normalize{$Q$} \;

\BlankLine

\tcp{Selecting candidates corpus examples}

\ForEach{$(M_{node}, w_{node}) \in Q$}{

    $D_{f} \gets$ \Retrieve{$M_{node}, D$} \;

    $C \gets$ \Merge{$C, D_{f}, \text{init\_score} \gets 0$} \;

}

\BlankLine

\tcp{Computing scores for corpus examples}

\ForEach{$c \in C$}{
    \ForEach{$(M_{node}, w_{node}) \in Q$}{
        $hits \gets$ \CountHits{$c.M, M_{node}$} \;
        $c.score \gets c.score + (hits \times w_{node})$ \;
    }
}

\BlankLine
\tcp{Selecting K corpus examples}
$R \gets$ \Sort{$C$, by=$c.score$, order=$\downarrow$} \; 
$R \gets R[0:K]$\;

\Return{$R$}
\end{algorithm}

\subsubsection{Retrieval}

\label{subsec:retrival}
To transition from the logical blueprint to a concrete implementation, selecting an effective modules matching strategy is crucial. We consider two approaches: a standard LLM-driven semantic retrieval method (as a baseline) and our proposed Structure-Aware Modules Matching approach.

In the LLM-driven baseline, retrieval is performed by directly prompting the model to select relevant code summaries based solely on semantic reasoning (textual similarity).

In contrast, our proposed approach explicitly leverages the Logical Pipeline Plan constructed in the previous stage. We observe that the correctness of a visualization pipeline depends strictly on using the correct operators (e.g., a ``Slice'' node must be implemented by \texttt{vtkImageSlice}). Since the Candidate Modules for each node have already been predicted by the Architect Agent, we utilize these module names as structural anchors to retrieve validated pipeline fragments. This ensures that the retrieved code is not just semantically relevant to the description, but functionally compatible with the planned pipeline topology.

\textcolor{review}{Algorithm~\ref{alg:vtksercher_v2_simple} illustrates structure-aware matching and re-ranking procedure. The input ${D}$ is a list of corpus entries, where each entry is represented as a triple $(M,\text{code},\text{des})$. Here, $M$ denotes the set of vtk.js modules used in the corresponding code snippet, and \texttt{code} and \texttt{des} represent the code snippet and its associated functionality, respectively. $Q$ represents a Pipeline Plan, we formalize it into a list of pipeline node (each node represents one step in the pipeline). In each node, $M_{node}$ represents vtk.js modules associated with corresponding step, and $w_{node}$ represents the weight (importance value) of current node in the context of the corresponding pipeline.}

\textcolor{review}{The initialization of data structures in lines 1-2 corresponds to auxiliary data structures initialized within the algorithm.
With these variables in place, Lines 3–5 perform the core retrieval step: for each Pipeline Node, the function \texttt{RetrieveByModules} matches the modules $M_{node}$ against the corpus to identify code examples containing the required modules, forming the candidate set $C$.
From lines 6–9, for each candidate $c \in C$, the function \texttt{CountHits} calculates the overlap between the candidate's actual modules $c.M$ and the node's predicted modules $M_{node}$. The score is updated by $hits \times weight$, ensuring that fragments matching critical topological nodes (high weight) are prioritized over those matching auxiliary steps. Finally, the top-$K$ pipeline fragments are returned for assembly.}

\subsubsection{Generation}

After the retrieval stage, the system obtains a set of corpus entries that are relevant to the current query. These retrieved items, together with the original query, are then assembled into a structured prompt for code generation. Since large language models are prone to hallucination, the prompt must impose clear and rigorous constraints to ensure the generation of valid and expected outputs. Such constraints include the required response format, coding style, and the specific vtk.js version to be used.

For the proof of concept, the generator is required to output a complete, self-contained HTML document beginning which can be validated through browser conveniently. All reasoning performed by the model is restricted to code comments to prevent unintended natural-language output from polluting the final result. The prompt further enforces several constraints to ensure reliability: (1) the generated content must consist solely of executable HTML and JavaScript code; (2) any explanation must appear only as in-code comments; (3) the generator must strictly follow the usage patterns of vtk.js without fabricating nonexistent APIs; and (4) when relevant corpus examples are provided, the model is instructed to adapt and extend these examples rather than producing code from scratch. These constraints collectively help mitigate hallucination, enforce syntactic correctness, and ensure that the produced visualization pipeline adheres to the semantics implied by the retrieved examples.

Even with carefully designed prompts, different LLMs can generate distinct outputs due to varying generation capabilities. To assess these differences, we employ multiple approaches compare the quality of their generated results. The generated visualizations and associated code require detailed evaluation, and the methodology for this evaluation is described in the following section.

\subsection{Evaluation principles}
\label{method:principles}

In code-generation tasks for scientific visualization, the correctness of the \textcolor{review}{generated code is critical.} A model may generate code that is largely correct, yet a minor error—such as an incorrect rendering parameter—can prevent any visualization from being displayed, making it impossible to evaluate the result purely from its visual output. To enable more objective assessment under such conditions, we introduce a set of multi-faceted evaluation criteria and a corresponding scoring scheme tailored to visualization-oriented code generation. In the following, we describe these scoring rules in detail and explain how we construct the test dataset used for evaluation.

\subsubsection{Grading strategies}
\label{sec:grading}

To ensure a consistent and comprehensive assessment, we established a unified evaluation framework utilized by both automated agents and human experts. This framework encompasses three fundamental dimensions. First, \textit{Functionality and Completeness} verifies that the code faithfully implements the user's semantic intent and preserves essential pipeline modules—such as filters and mappers—rather than merely checking for execution errors. Second, \textit{Visual Fidelity} assesses the effectiveness of rendering parameters (e.g., camera views, color transfer functions, opacity) in revealing scientific data patterns and their alignment with ground truth. Finally, \textit{Code Quality and Maintainability} examines the script's clarity, structure, and adherence to \textit{vtk.js} best practices to ensure support for real-world research workflows.Guided by this framework, we implemented a two-stage evaluation process. The workflow begins with an LLM-based automated evaluation, where an LLM evaluator analyzes the generated code against the aforementioned dimensions. It assigns a numerical score in the range [0.0, 1.0] for each metric and provides reasoning, serving as a rapid and scalable quality check. Subsequently, a Human-in-the-Loop evaluation provides a more rigorous validation. 
We introduce a primary metric termed correction cost, which quantifies the manual effort (lines added, removed, or modified) required to restore the pipeline's topological validity. The guiding principle for correction is to ensure that every individual pipeline node functions correctly and strictly adheres to the intended data flow constraints. Upon successful rendering, experts further perform a qualitative assessment across the same three fundamental dimensions to complete the evaluation.

\subsubsection{Ground-Truth Case Construction}
\label{sec:bench}

To support the evaluation of generated visualization code, we construct a set of ground-truth cases adapted from vtk.js examples. These cases serve as reference implementations against which the generated results can be compared. The collection spans three major categories—I/O, Filter, and Rendering—covering a representative range of visualization operations. Each case consists of a description.txt file specifying the intended behavior and a corresponding code.html implementation that fulfills the specification. The HTML files load vtk.js through the unpkg CDN, ensuring easy execution, dependency-free validation, and reproducible rendering across environments. Most cases follow a standard Source–Filter–Rendering pipeline, while several extended cases include multiple filtering stages. Although the number of cases is modest, they provide a controlled and reliable set of ground-truth references for evaluating correctness and computing correction cost. Experiments section detail how these ground-truth cases are used to quantify correction cost across different code-generation results for visualization tasks.

\subsection{Interactive Visualization System}
\label{method:vis}


\begin{figure*}[htbp]
\centering
\includegraphics[width=\textwidth]{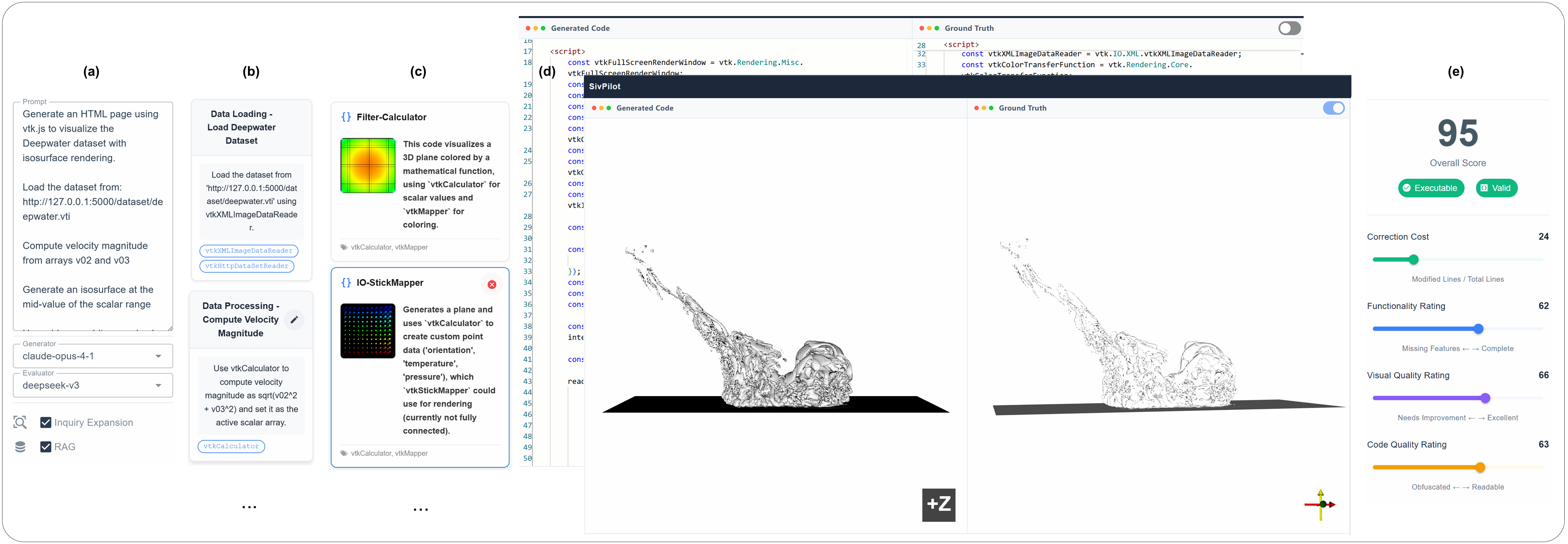}
\caption{The interactive visualization system supports end-to-end construction and verification of scientific visualization pipelines. The interface is organized into three regions: the left panel (a–c) implements the multi-stage augmentation workflow, the middle panel (d) provides synchronized views for code editing and rendering inspection, and the right panel (e) supports multi-dimensional assessment. In particular, (a) Configuration region allows users to enter natural-language instructions and select target models. (b) Pipeline Planning region visualizes the decomposed, pipeline-aligned steps, enabling users to refine intermediate logic before generation. (c) Corpus Retrieval presents retrieved reference cases with functional summaries and visual previews, and includes a rejection mechanism to filter irrelevant context. (d) Rendering and Editing offers a dual-view workspace to debug generated code by comparing its rendered output with the expected result. (e) Evaluation Panel reports automated scores and enables human-in-the-loop qualitative assessment across multiple criteria.}
\label{fg:system}
\end{figure*}

To validate the effectiveness of the proposed workflow and support ``Human-in-the-loop'' assessment, we developed a prototype system designed to move beyond traditional ``black-box'' code generation. As illustrated in Figure \ref{fg:system}, the interface is organized into three primary functional areas: the left panel (a-c) serves as the augmented processing stack; the middle panel (d) provides a toggleable environment for code editing and real-time rendering; and the right panel (e) facilitates multi-dimensional evaluation. We stack (a–c) within the left panel to provide an overview of the multi-stage augmentation workflow; in the actual system, (a), (b), and (c) are separate pages, and users move between them as they progress through the workflow.

The workflow begins with the expansion of the user's initial intent into a structured visualization pipeline. By decomposing ambiguous natural language instructions into concrete pipeline nodes, the system bridges the gap between high-level user intent and the rigid procedural logic required by scientific visualization frameworks. These steps are visualized as interactive nodes, letting users verify the planned execution sequence and manually refine the logical flow if its decomposition deviates from specific requirements.

Following the refinement of the logical chain, the system executes a module-aware retrieval algorithm to fetch relevant reference code from the curated corpus. This stage provides the downstream generator with explicit implementation grounding, which is critical for reducing the risk of API hallucinations in domain-specific environments like vtk.js. To assist users in evaluating the suitability of the retrieved evidence, each code snippet is presented alongside functional descriptions and visual thumbnails. Through an integrated ``Rejection Mechanism'', users can filter out irrelevant context, ensuring that only high-quality reference patterns are used for the final code synthesis.

Once the implementation code is generated, the system facilitates an iterative debugging and adjustment process through its dual-view editor and real-time visualization window. This layout allows for direct comparison between the generated results and the ground-truth rendering, providing immediate visual feedback on the pipeline's correctness. Furthermore, the system captures runtime logs via a console panel; in the event of execution failures, it automatically parses these logs to highlight the problematic lines of code, thereby assisting users in performing precise manual corrections.

The final stage of the workflow completes the loop through a comprehensive, multi-tiered assessment process. The system first performs automated verification, encompassing syntax checking and execution monitoring to ensure code viability. This is followed by an automated evaluation performed by an LLM-based agent, which analyzes the generated code against three fundamental dimensions: functionality, visual fidelity, and code quality. As detailed in the evaluation principles, this assessment is grounded in a systematic comparison against pre-constructed ground-truth acting as high-quality reference implementations. This comparison allows the system to quantify the generation's reliability through metrics such as correction cost, which measures the manual effort required to align the generated code with the ground truth for successful execution. Finally, human experts conduct a rigorous qualitative evaluation using the interactive scoring panel, adjusting the multi-dimensional ratings to provide final validation of the pipeline's adherence to scientific standards.

\begin{figure*}[ht]
    \centering
    \includegraphics[width=1.0\textwidth]{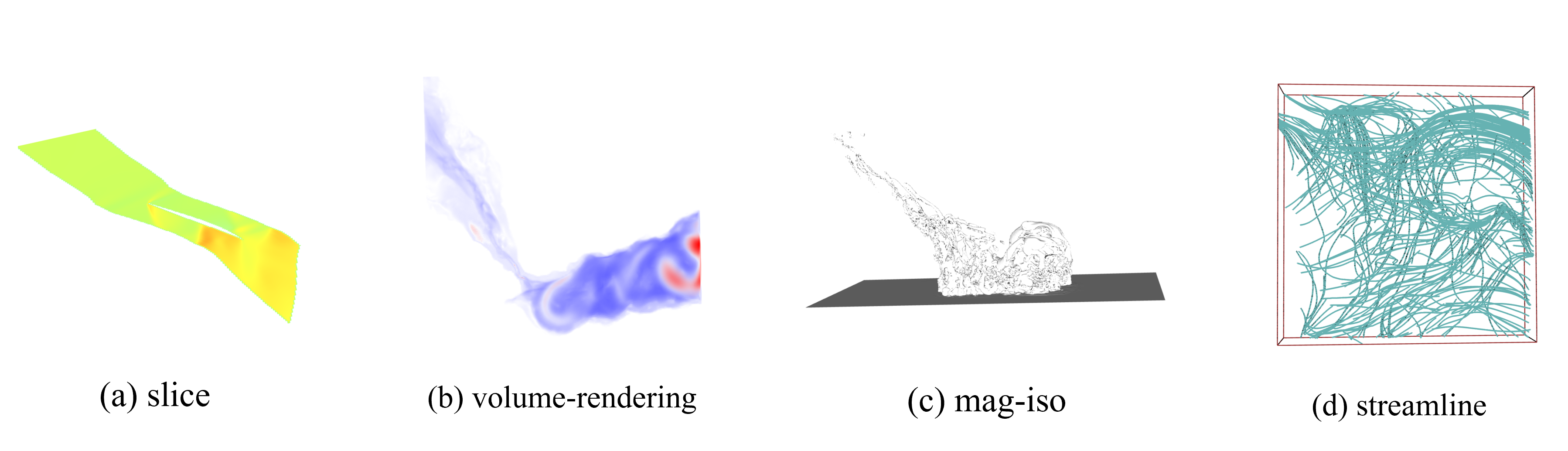}
    \caption{Overview of representative visualization tasks on the four evaluated datasets. (a) Slice-based exploration of the Rotor dataset using a slice filter. (b) Streamline visualization of the velocity field using the Isabel dataset. (c) Mag-iso extracts isosurface of the magnitude variable derived from arrays. (d) Volume rendering results for the RedSea dataset.}
    \label{fig:exp:visall}
\end{figure*}

\section{Experimental Results}

This section provides a comprehensive evaluation of the proposed approach. We first introduce the experimental data and computing platform, then present the evaluation of the vtk.js example retrieval component. After that, we assess the effectiveness of the RAG workflow introduced in this work. Finally, we report the results of applying the RAG workflow across different code-generation models.

\subsection{Evaluated datasets and LLMs}
\label{exp:data_platform}
This section presents details of the datasets and the LLMs used during evaluation.

\textbf{Rotor} is derived from a steady-state simulation of the NASA Rotor 37~\cite{denton1997lessons}. This computational fluid dynamics data captures the complex flow patterns within the compressor geometry. Figure~\ref{fig:exp:visall}(a) visualizes a pressure field slice through the rotor assembly, highlighting the aerodynamic characteristics and pressure distribution.

\textbf{DeepWater}~\cite{patchett_deep_water_2017} simulates asteroid impacts on ocean surfaces, modeling the complex fluid dynamics. This dataset captures various impact scenarios with different parameters. Figure~\ref{fig:exp:visall}(b) illustrates isosurface of key physical variables, demonstrating the propagation of impact waves and energy distribution in the water column.

\textbf{Isabel} is a large-scale multivariate atmospheric simulation released as the contest data set for IEEE VIS 2004~\cite{isabel2004}. Figure~\ref{fig:exp:visall}(c) shows streamlines generated from Isabel's velocity field, providing a visual representation of flow patterns within the dataset.
 
\textbf{RedSea}~\cite{kaust_scivis_2020} is released by 2020 SciVis Contest. The data set shows oceanology simulations with varying parameters. Figure~\ref{fig:exp:visall}(d) illustrate the volume for this data set with depth. 

\textcolor{review}{We select four representative scientific visualization use-case scenarios based on the datasets described above. These cases correspond to common and widely adopted visualization representations in scientific data analysis, and are frequently used in prior work on visualization-oriented code generation task using Python scripts~\cite{chatvis}. The resulting visualizations are shown as proof-of-concept examples in Figure~\ref{fig:exp:visall}.}

We employ different LLMs according to the goal of the experiments. The selection of LLMs is based on their established strengths and prevalence in code-related tasks. For the pipeline planning, we employ Qwen-Turbo \cite{yang2025qwen3} for its robust instruction-following and reasoning capabilities. During code generation, we use three representative, state-of-the-art models: GPT-5~\cite{singh2025openai}, Claude-4-Sonnet~\cite{Anthropic2026ClaudeSonnet}, and DeepSeek-V3~\cite{liu2024deepseek}, all of which are widely recognized and benchmarked for their superior performance in code synthesis. Finally, for the evaluation stage, we leverage DeepSeek-R1~\cite{guo2025deepseek}, a model specifically designed for reasoning tasks and widely adopted. It is worth noting that although we use LLMs in the evaluation above, the methodology presented in this work is not constrained to these particular LLMs.

\subsection{Code corpus retrieval}
\label{exp:retrieval}

This section presents the experimental methodology, evaluation metrics, and results for the code corpus retrieval task. Given a user-specified query, the goal is to identify the most relevant code snippets from a large corpus in order to support downstream code generation. In our experiments, we compare two representative retrieval paradigms: direct semantic retrieval using LLM (as a baseline) and structure-aware module matching (our proposed method).

\textcolor{review}{The corpus used in the evaluation contain 30 carefully selected, representative examples drawn from the official vtk.js examples website. Each example consists of three files: \texttt{code.html}, which contains the original vtk.js example; \texttt{description.txt}, describing the general functionality of this code; and \texttt{meta.json}, used for indexing specific example.}

To systematically evaluate retrieval performance, we designed four representative code-retrieval tasks, covering common visualization operations such as data slicing, volume rendering, derived-field computation and isosurface extraction, and streamline generation. These tasks are derived from practical requirements in web-based scientific visualization applications. The specific task definitions are described as follows:

\begin{itemize}
  \item \textbf{Slice:}
  Generate an HTML page using vtk.js to visualize the rotor dataset: load the dataset from [data server]; set the active scalar array to ``Pressure''; apply a slice along the Y axis at 95\% depth of the dataset (convert percentage to slice index); use a blue $\to$ white $\to$ red color map for pressure values, spanning from the minimum to maximum scalar range; set opacity to fully opaque.
  
  \item \textbf{Volume rendering:}
  Generate an HTML page using vtk.js to visualize the Redsea dataset with volume rendering: load the dataset from [data server]; compute velocity magnitude from the ``velocity'' array and set it as the active scalar; apply volume rendering using a blue $\to$ white $\to$ red color map spanning the scalar range; apply a piecewise opacity function.

  \item \textbf{Mag-Iso:}
  Generate an HTML page using vtk.js to visualize the Deepwater dataset with isosurface rendering: load the dataset from [data server]; compute magnitude from arrays [field names]; generate an isosurface at the mid-value of the scalar range; use a blue $\to$ white $\to$ red color map spanning the scalar range.

  \item \textbf{Streamline:}
  Generate an HTML page using vtk.js to visualize the Isabel dataset with streamline rendering: load the dataset from [data server]; use the ``Velocity'' array as the vector field for streamlines; generate seed points at the center of the dataset with sufficient density to cover the domain; compute streamlines following the velocity field.

\end{itemize}

Across the four visualization tasks, two evaluated retrieval-corpus strategies exhibit strong functional coverage by consistently recalling the essential vtk.js building blocks required to assemble the core visualization pipelines. Specifically, both methods retrieve the critical modules for each task—e.g., volume-mapper and volume-clip code example for volume rendering, streamline code example for streamline extraction, code containing slice example is selected for slice-based rendering, and code containing marching cube is selected for isosurface generation. These results suggest that, regardless of corpus construction, retrieval can reliably provide the key operators needed for pipeline composition, and the remaining performance differences are more likely attributable to factors such as module ranking, redundancy, or the correctness of subsequent pipeline integration and parameterization.

To further evaluate retrieval effectiveness, we measured the runtime of the two retrieval strategies. The module-aware retrieval by module matching is highly efficient, achieving an average latency of around 0.01 s, while the LLM-based direct semantic retrieval requires approximately 17s on average (DeepSeek-V3). Combined with the observation that both strategies consistently recall the essential vtk.js modules across the four visualization tasks, these results indicate that module matching can deliver comparable retrieval coverage and downstream pipeline capability with orders-of-magnitude lower computational cost, making it more suitable for interactive or rapid prototyping scenarios.

Since our experiments require performing large numbers of retrieval operations across multiple tasks, and it is essential to maintain high recall and retrieval quality while also ensuring overall efficiency, we adopt the module-matching retrieval strategy as the default method for obtaining contextual information in all subsequent experiments.

\subsection{Effectiveness of RAG workflow for Scientific Visualization Pipeline Construction}
\label{exp:rageval}

\begin{figure}[H]
    \centering 
    \includegraphics[width=0.48\textwidth]{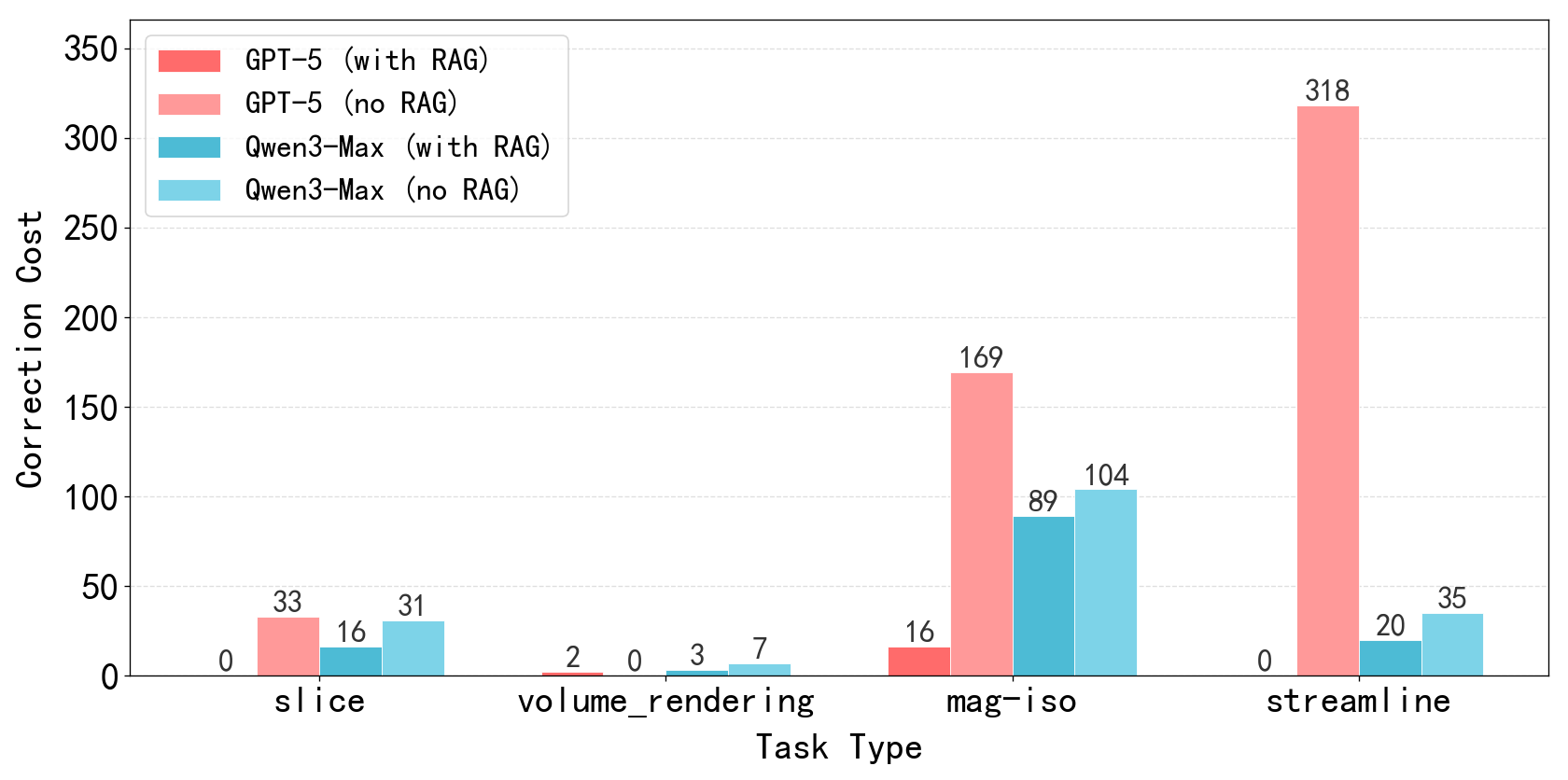}  
    \caption{Comparison of correction costs for scientific visualization tasks with and without the RAG-based workflow.}
    \label{fig:correction_cost_is_rag}
\end{figure}

Figure~\ref{fig:correction_cost_is_rag} compares the correction cost using two state-of-the art LLMs under two conditions: with and without the RAG-based workflow. Correction cost is computed using \texttt{difflib} by counting the total number of added and deleted lines required to transform the generated code into the corrected reference version.

Overall, the results show that incorporating RAG-based workflow yields a substantial reduction in correction cost across all tasks. For GPT-5, the improvement is especially significant: when using RAG-based workflow, GPT-5 produces code that is either immediately executable or requires only minimal adjustments. In the slice, streamline, and volume rendering tasks, the generated outputs closely match the reference implementation and can often be rendered without any manual modification. 

For streamline generation, GPT-5 tends to attempt a full implementation “from scratch” when no RAG support is provided, leading to complex and unstable code that cannot be executed directly. The RAG-based workflow mitigates this behavior by providing concrete examples, enabling GPT-5 to use \texttt{vtkImageStreamline} filter to get expected results quickly.

A similar trend is observed for Qwen3-Max, although the improvements are less pronounced. With RAG-based workflow, Qwen3-Max shows meaningful reductions in correction cost, yet still requires noticeable manual effort, particularly for isosurface and streamline. A closer inspection reveals the cause: Qwen3-Max frequently relies on the vtkCalculator filter to compute derived scalar fields. This operator is highly error-prone because it demands strict configuration of input arrays, output array names, and expression syntax. As a result, large portions of the generated code must be rewritten during correction. 

Taken together, these results demonstrate that RAG-based workflow significantly improves the reliability and usability of generated visualization code, with the strongest gains observed for GPT-5. While Qwen3-Max also benefits from RAG-based workflow, its reliance on brittle operators such as vtkCalculator and its difficulty handling complex pipeline stages indicate that further model-specific adaptation may be required.

\subsection{Correction cost across different models}
\label{sec:exp_different_models}

\begin{figure}[H]
    \centering
    \includegraphics[width=0.48\textwidth]{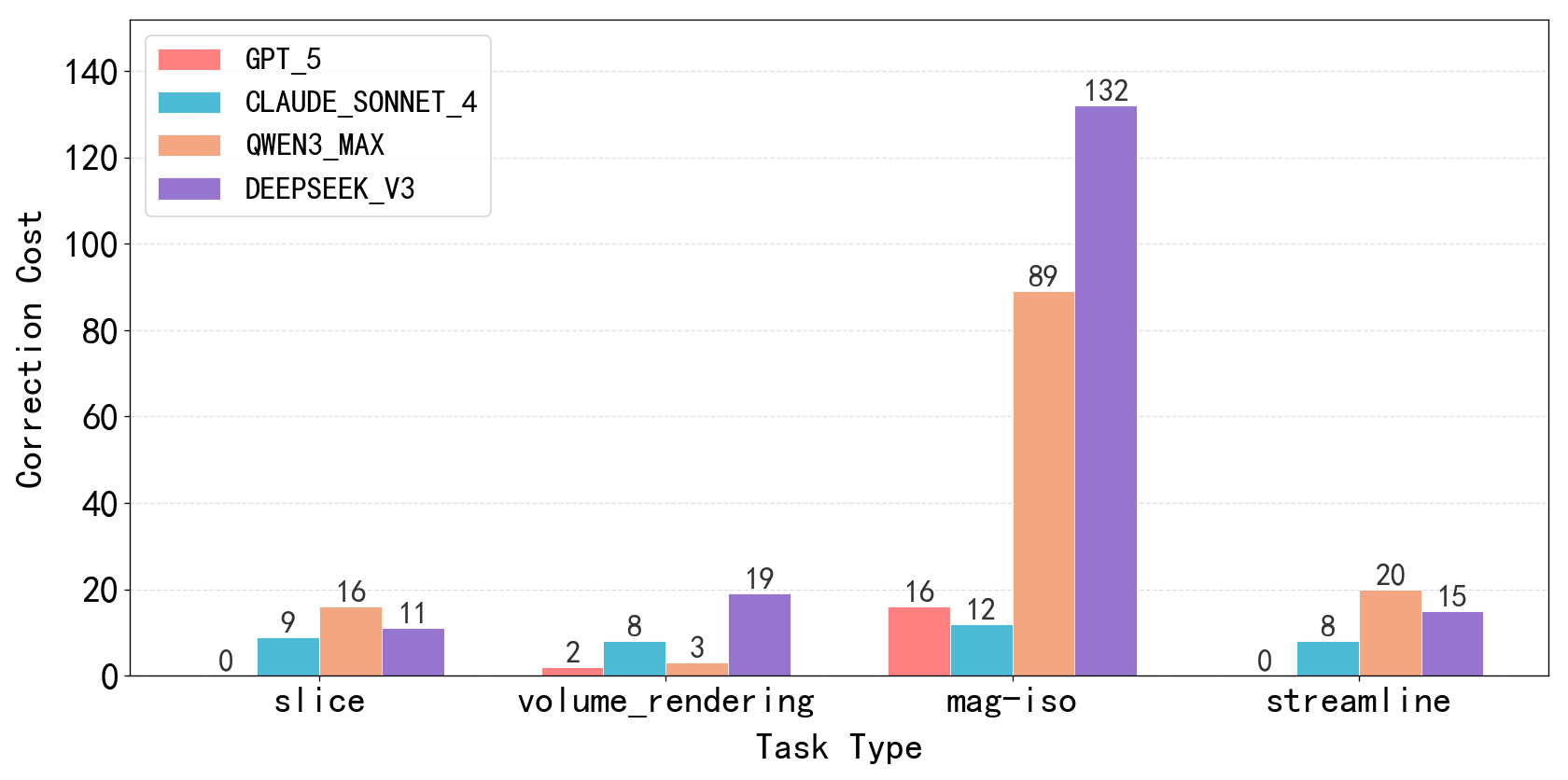}  
    \caption{Correction cost using RAG-based workflow with different LLMs}
    \label{fig:correction_cost_models}
\end{figure}

Figure~\ref{fig:correction_cost_models} summarizes the correction cost across four representative visualization tasks. Several observations can be made from the results.

First, although all models operate under the same RAG workflow through RAG-based workflow, their performance varies substantially, indicating that retrieval-augmented generation alone does not eliminate differences in model capability. Among all evaluated systems, GPT-5 consistently achieves the lowest correction cost across all tasks, demonstrating the strongest robustness in producing syntactically correct and semantically meaningful vtk.js code.

Second, magnitude-isosurface extraction exhibits notably higher correction costs for Qwen3-Max and DeepSeek-V3. A closer inspection reveals that both models tend to employ the vtkCalculator filter to compute derived scalar arrays instead of implementing the operation directly with JavaScript loops. While this choice is conceptually sound, it is highly error-prone in practice because vtkCalculator requires strict configuration of input arrays, result array names, and data types. As a consequence, large portions of the generated code must be rewritten during correction. If errors introduced by the calculator filter are excluded, the correction cost for Qwen-Max and DeepSeek-V3 becomes comparable to that observed in the volume-rendering task.

For volume rendering task, the overall correction cost remains low across most models, indicating strong end-to-end stability for this task. GPT-5 and Qwen3 Max require only minimal edits, while Claude Sonnet 4 also stays relatively low. DeepSeek V3 shows a higher correction cost. After carefully inspecting the results, we find that error mainly caused by computing the velocity magnitude instead of rendering settings for DeepSeek V3.

The mag-iso task is the most challenging among the four: in addition to the data-processing steps, it requires configuring multiple rendering parameters, which models often struggle to specify correctly using natural-language supervision alone. These results highlight both the strengths and current limitations of state-of-the-art LLMs in generating reliable visualization pipelines, even when supported by the proposed RAG-based workflow.

\section{Discussion}

Our study highlights both the promise and current limitations of LLM-assisted visualization pipeline construction in vtk.js. Across the evaluated models and tasks, retrieval-augmented generation improves end-to-end executability and reduces manual repair effort, but does not fully eliminate failures. To better understand where errors arise and how our workflow helps, we first summarize recurring \emph{error types} observed during manual correction, then attribute them to three underlying \emph{reasons} that connect directly to our experimental results. Finally, we discuss practical implications and limitation in this study.

\paragraph{Error type analysis}
After comparing the original generated code with our manually fixed versions and reviewing the detailed repair process, we found that errors in LLM-generated visualization pipelines can be grouped into four recurring types, roughly ordered from easier to harder to fix:

\begin{enumerate}
    \item \textbf{Module and import issues} These are the most frequent errors and typically easy to repair, including missing or redundant module declarations, mixing incompatible packages (e.g., \texttt{vtk} vs. \texttt{vtk.js}-specific modules), and incorrect import paths.
    \item \textbf{Missing or incomplete parameterization} These errors often break executability or lead to incorrect results, such as failing to set active scalars/vectors, omitting key rendering configurations, or neglecting pipeline parameters required by downstream stages.
    \item \textbf{API hallucinations} These are more disruptive cases where the model invokes non-existent functions, uses invalid method names, or supplies parameters with incorrect types errors.
    \item \textbf{Wrong intention/logic} These are the hardest to fix, as the generated code follows an incorrect construction strategy (e.g., reimplementing streamline integration from scratch), requiring substantial rewrites rather than localized edits.
\end{enumerate}

We observe that query-based expansion and more structured contextual guidance through RAG can reduce wrong intention/logic errors and, to some extent, API hallucinations. Crucially, the accuracy and coverage of the underlying corpus have a first-order impact on generation quality. However, ensuring fully executable end-to-end pipelines still requires human verification through interactive visualization system. When pipeline-aligned context is available, the remaining human effort is typically limited to minor, localized corrections rather than large-scale rewrites.

\paragraph{Error reason analysis}
We further attribute the above error types to three underlying reasons, each of which is reflected in our experimental observations.

\begin{enumerate}
\item \textbf{Knowledge/Context gaps and retrieval alignment}: Many failures originate from missing vtk.js-specific knowledge (module conventions, required parameter settings). Our evaluation shows that both corpus strategies consistently recall the essential building blocks for the four tasks, suggesting that high-level operator recall is largely achievable. Meanwhile, the module-matching strategy achieves orders-of-magnitude lower latency, making it suitable for interactive workflows. These results indicate that remaining failures are less about whether key operators exist in the context, and more about whether the retrieved evidence is well-aligned with the intended pipeline stage and integrated together in a correct manner (e.g., correct module selection and parameter wiring).

\item \textbf{Intent-pipeline misalignment}: Strategy-level errors occur when the user intent is not translated into an appropriate pipeline decomposition, resulting in wrong intention/logic failures that often require non-local rewrites. This is most evident in streamline generation: without RAG support, models may attempt to implement streamline integration ``from scratch'', producing overly complex code that is difficult to debug and repair. In contrast, pipeline-aligned RAG steers the model toward established operators (e.g., \texttt{vtkImageStreamline}), enabling executable solutions with substantially fewer edits, consistent with the observed reduction in correction cost under the RAG-based workflow. Nevertheless, intent misalignment can still manifest as suboptimal operator choices: when the model fails to recognize the simplest intended computation, it may rely on brittle, template-driven components (e.g., using \texttt{vtkCalculator} for a straightforward magnitude calculation), which increases configuration overhead and introduces additional API using errors.

\item \textbf{API semantic complexity}:
Even with relevant context, intricate or template-driven APIs increase the likelihood of subtle mistakes, amplifying correction cost. This pattern is clearly reflected in isosurface extraction: Qwen3-Max (and similarly DeepSeek-V3) frequently relies on \texttt{vtkCalculator} to compute derived scalar fields, which requires strict configuration of input arrays, output names, and expression syntax. As a result, small specification errors can cascade into non-executable code and often necessitate rewriting larger code blocks during correction, yielding notably higher correction costs. These observations suggest that reliable pipeline generation depends not only on retrieval but also on API design. Well-encapsulated APIs that expose composable, high-level primitives with clear defaults and validated parameter schemas can reduce the opportunity for configuration errors, making both LLM generation and human repair more reliable. 

\end{enumerate}

\paragraph{Implications and Limitations}

Although RAG reduces hallucination in both retrieval and generation, it does not eliminate it; human oversight is still required. In practice, extracting key API modules during query expansion proves to be an effective way to enable efficient retrieval in visualization oriented code generation. Besides, many visualization tasks demand iterative adjustments, such as parameter tuning or colormap configuration, so an interactive visualization system is still essential.

Although the correction cost quantifies editing effort, it depends on how differences are computed. A single-line modification is counted as both a deletion and an insertion, and even changes in comments can increase the measured cost. Refactoring larger code blocks amplifies this effect. Despite these limitations, the metric still reflects human effort to a meaningful extent: more required edits generally correspond to more substantial corrections. \textcolor{review}{One potential solution to further improve objectivity is to explicitly distinguish visualization-related edits from refactoring or comment-level changes. Rather than counting all code differences uniformly, the correction cost could be computed only from modifications that directly affect visualization behavior, such as changes to visualization APIs, pipeline structure, or rendering and visual-encoding parameters.}

\textcolor{review}{In addition, the evaluated datasets are relatively limited and mainly consist of standard scientific visualization use cases.
Although different datasets are adopted for evaluated visualization use cases, the datasets for each individual use case are fixed.
Investigating how dataset variations across visualization scenarios affect the results could provide further insights into the generalization of our approach.}

\textcolor{review}{In practice, scientific visualization commonly involves large-scale volumetric datasets and intensive user interaction, such as camera manipulation and transfer function design \cite{bernardon2007interactive,sun2024rmdncache,ai2025nli4volvis}. How LLM-based code generation can effectively support such interactive and scalable visualization scenarios is, therefore, an important open question. We include proof-of-concept experiments in the supplementary materials to explore this setting. The results show that once a visualization pipeline is generated correctly, constructing interactive panels and controls is relatively straightforward. However, generating appropriate configurations, particularly transfer functions and parameter settings that faithfully matching user intent, remains challenging for current LLMs, especially in large-scale and web-based volume visualization systems. Addressing these challenges and extending reliability-oriented pipeline generation toward scalable rendering and efficient human–AI interaction represents an important direction for future research.}

\textcolor{review}{While our evaluation framework rigorously assesses the executability of the visualization pipeline, it currently lacks a quantitative metric for fine-grained visual quality. In our present workflow, visual fidelity (e.g., the correctness of color mapping configurations and rendering parameters) is primarily verified through human-in-the-loop inspection to ensure alignment with ground truth. We acknowledge that manual verification is not scalable for large-scale benchmarks. To address this, we plan to integrate Multimodal Large Language Models (mLLMs) into our evaluation loop in future work. By leveraging these models to automate comparisons between generated renderings and ground truth using perceptual similarity and semantic consistency, we aim to establish a more comprehensive, automated scoring system for visual quality.}

\section{Conclusion and Future work}

We investigated the reliability of constructing scientific visualization pipelines from natural-language descriptions using vtk.js.
Unstructured LLM generation frequently fails to produce executable pipelines due to missing stages, incompatible operators, or incorrect execution order.
Conversely, a retrieval-augmented workflow exposing models to pipeline-aligned, structure-aware examples substantially improves end-to-end executability and reduces human correction effort.

Our findings highlight that reliable pipeline generation requires both language modeling capacity and explicit structural and semantic cues at inference time.
Concretely, combining a curated domain corpus, compatibility-aware retrieval, and constrained prompting enables LLMs to better respect stage ordering, operator selection, and parameterization requirements.
By introducing correction cost as a reliability metric, we provide a more informative measure of the human effort required to repair non-executable pipelines.

We further showed that interactive support for execution-based validation plays an important role in evaluation and error analysis.
Allowing users to inspect retrieved evidence, execute pipelines in-browser, and apply targeted corrections facilitates transparent human-in-the-loop assessment of generation behavior and failure modes.

Despite these improvements, reliable pipeline generation remains challenging for visualization tasks involving complex semantic dependencies, derived-field computation, or tightly constrained parameter spaces.

Future work will explore adaptive retrieval strategies that account for inter-stage dependencies, alongside mechanisms for automatically synthesizing configurable UI components and camera settings.

\textcolor{review}{Additional scientific visualization code generation tasks beyond user case scenarios discussed in this paper remain to be explored, including other popular web-based visualization frameworks such as three.js~\cite{threejs_manual}.}
More broadly, we see opportunities to extend reliability-oriented evaluation to other domain-specific visualization frameworks, and to further study how structured context and human feedback can jointly support trustworthy visualization authoring with large language models.

\section*{Data and Code Availability}
The source code, curated datasets, and the interactive visualization system introduced in this study are available on GitHub at \url{https://github.com/Indigo-gg/llmscivis}.

\begin{acks}
This paper will be published in Information Visualization Journal. We used ChatGPT 4o and Gemini solely for grammar correction and language polishing.

\end{acks}





\end{document}